\documentstyle[12pt]{article}
\title{Selfcoupled equations for  the field correlators}
\author{Yu.A.Simonov\\ Institute of Theoretical and
Experimental Physics\\ 117218, Moscow, B.Cheremushkinskaya 25,
Russia}
 \date{}
\newcommand{\be}{\begin{equation}}
\newcommand{\ee}{\end{equation}}

\begin{document}
\maketitle

\begin{abstract}

New  nonlinear equations for field correlators in the large $N_c$
gluodynamics are  derived. Nonperturbative confining correlators
 are shown to satisfy a system of  nonlinear scale--invariant
 equations.  Keeping only the simplest,  bilocal correlators one can
find  the exponential large distance behaviour for both independent
correlators $D$ and $D_1$ and express  coefficients in the
exponentials  -- the gluonic correlation length $T_g$ -- through  the
string tension.  Numerically these results are in agreement with
lattice data.

\end{abstract}

\section{Introduction}

The main unsolved problem of QCD and of  the modern field theory in
general is the quantitative treatment of nonperturbative phenomena.
There is no  formalism yet (e.g. a system of equations) in
which nonperturbative quantities, like masses and condensates in QCD, are
computed from the first principles. By now there are two main
approaches to this problem: i) lattice calculations; ii) modelling the
nonperturbative vacuum with e.g. quasiclassical solutions.
In the first approach one can indeed calculate numerically hadron masses and
even more fundamental quantities like field correlators [1]. However,
here the physical mechanism creating nonperturbative phenomena is
kept hidden behind the formidable numerics.
The second approach is necessarily model dependent and approximate.
 Also there is no successful model  of the QCD vacuum, which
describes both basic nonperturbative phenomena: confinement and
chiral symmetry breaking (CSB) (however,  the dyonic model recently
suggested for QCD [2]  might be promising in this respect).

The main characteristics of nonperturbative (NP) phenomena in QCD is that
they create there own mass scale. By this fact the scale invariance
is violated and through the scale anomaly the NP shift
downwards in the vacuum energy density   is produced [3].
Therefore one may expect that NP   solutions appears  as
scale violating solutions of some nonlinear scale--invariant
equations, while  perturbative solutions of the same equations are  obtained
by expanding in powers of coupling constant.

Recently equations of this type  have been written for
gauge--invariant quark propagators [4], where the  kernel of the
equations contains the fundamental string.
The equations [4] do not violate symmetry (in this case the chiral symmetry),
while NP solutions break chiral symmetry.
 In [4]  the gluon-field
correlators were taken as an input and
 the problem how to find the correlators from the first principles
was not solved there. Here we are proposing   selfcoupled equations
for field correlators in  the  case of gluodynamics. These equations are
also of Dyson--Schwinger type but will be written for the gauge--invariant
amplitudes, which automatically display at large distances the formation of
adjoint string in the large $N_c$ limit.

 To write these equations we divide all
$N_c^2-1$ gluonic fields into two  groups: a small group of $r$
fields $b^i_{\mu}, i=1,...r$, which later be called "valence gluons",
and a large group of fields $A^a_{\mu}$, $a= r+1, ... N^2_c-1,$ which
 form a background for valence gluons.

As will be shown below this  background   creates in the  confining
phase after vacuum averaging an adjoint string. Hence, physically the problem
reduces to the motion of valence gluons in the  field of the string.  Since
the string is a white object, the action of the  string on valence gluons
generates color--diagonal mass operator and the resulting equation for the
valence gluon propagator simplifies considerably.  We shall find out
 that these equations are nonlinear and selfcoupled, so that one can look
 for selfconsistent solutions for the valence gluon propagator and the
 field correlator.  Inserting the resulting field correlator in the kernel,
one realizes the selfcoupling  mentioned above.

Some properties of the  solutions  can be found even before the actual
solution of equations  e.g. the exponential large distance behaviour
of field correlators and
the coefficient in the exponent -- the so-called  gluon
correlation length $T_g$ -- can be computed through the  string tension
or gluonic condensate. Our number for $T_g$ is in agreement with
lattice calculations [1].

Thus one  finds a solution depending on a single scale. This scale
should be given as an  input to define the NP part of
the theory completely.

We shall not touch in this letter the
perturbative expansion and renormalization of equations obtained
here, referring  the reader to  a future  publication.
The paper is organized as follows. In section 2 the separation of
gluonic fields into "valence" and background parts is done and the
effective  action for the valence gluons is written explicitly. In
section 3 the mass operator for valence gluons is found after
averaging over background fields.
In section 4 the  connection of field correlators with the valence
gluon propagator is established and  it is used to  find the
asymptotic behaviour of field correlators and to compute the  gluonic
correlation length $T_g$ through the string tension or gluonic
condensate.
In Conclusion  a summary of results is given and possible
 extensions  are discussed.

\section{Effective action for valence gluons}

We start
dividing the gluonic fields
$A^a_{\mu}( a=1,... N_c^2-1)$ into a group of valence gluon fields
$b^a_{\mu}$, with the color index $a$ running a small set
of colours: $a=1,...r$,
while the rest of fields, for which we keep notation  $\bar A^a_{\mu}$,
belong to the large set, $a=r+1,... N_c^2-1$. Number of colors
$N^2_c-1$ is assumed to be large, and $r\ll N_c^2-1$.
Note that this division does not separate perturbative gluons from a NP
 background, as is usually done in the background perturbation method [5].
Both $b_{\mu}$ and $\bar A_{\mu}$ may contain perturbative and NP
components.  Hence, $A^a_{\mu}=\bar A^a_{\mu} + b^a_{\mu}$ and one
can write for the Euclidean action density
 $$ S_E(A)= S_E(\bar A+b)=
 $$ \be =S^{(0)}(\bar A)+ S^{(1)}(b,\bar A)+S^{(2,1)}(b,\bar
A)+S^{(2,0)}(b)+ \ee $$ S^{(2,2)}(b,\bar A)+S^{(3,1)}(b,\bar A)+
S^F(b,\bar A)+S^{(3)}(b)+S^{(4)}(b), $$
 where we have defined
 \be
S^{(0)}(\bar A)=\frac{1}{4}(F^a_{\mu\nu}(\bar A))^2,
 \ee
  \be
S^{(1)}(b,\bar A)=-b^i_{\nu}D^{ia}_{\mu}(\bar A) F^a_{\mu\nu}(\bar A)
 \ee
  \be
S^{(2,0)}(b) = b^i_{\mu}(-\frac{1}{2}\partial^2_{\lambda}
\delta_{\mu\nu}+\frac{1}{2} \partial_{\mu} \partial_{\nu}) b^i_{\nu}
 \ee
 \be
S^{(2,1)}(b, \bar A)=\frac{1}{2} gf^{ikc} b^i_{\mu}[2
\bar A^c_{\lambda}\partial_{\lambda}\delta_{\mu\nu} +
\stackrel{\leftarrow}{\partial}_\mu
\bar A^c_\nu-\bar A^c_\mu\partial_\nu] b^k_\nu
\ee
\be
S^{(F)}(b,\bar A)=gf^{ika}b^i_\mu b^k_\nu F^a_{\mu\nu}(\bar A)
\ee
\be
S^{(2,2)} (b,\bar A)=\frac{1}{2} \chi^{ik,ab}(b^i_{\mu} b^k_{\mu}
\bar A^a_{\lambda}
\bar A^b_{\lambda}-b_{\mu}^i b_{\nu}^k\bar A^a_\mu \bar A^b_\nu)
\ee
\be
S^{(3,1)} (b,\bar A)=
b^i_{\mu} b^k_{\mu}
\chi^{ik,ba}
b^b_{\nu} \bar A^a_{\nu}
\ee
\be
S^{(4)} (b)=\frac{1}{4}
b^i_{\mu} b^k_{\mu}
\chi^{ik,lm}b^l_{\nu} b^m_{\nu},~~
S^{(3)}=\frac{1}{2} gf^{ikl}b^i_\mu b^l_\lambda
\partial_{\lambda}b^k_{\mu}
\ee

In (2)-(9) the following notations are used
\be
\chi^{ik,ab}\equiv g^2\sum^{N^2_{c}-1}_{c=1} f^{cia}f^{ckb},
\ee
Note also that we have denoted color indices of $b_{\mu} $   fields by
$i,k,l,...$ running from 1 to $r$, and indices $a,b,c$ in
$\bar A_{\mu}$ running from $r+1$ to $N^2_c-1$.

As a next step consider averaging over fields $\bar A_{\mu}$ in some
expression, like the following
\be
D_{\mu\nu,\lambda\sigma}(x,y)=<F^\alpha_{\mu\nu}(x)\Phi^{\alpha\beta}(x,y)
F^\beta_{\lambda\sigma} (y)>_{b,\bar A},~~1\leq \alpha,\beta \leq N^2_c-1,
\ee
where $\Phi$ is the parallel transporter in the adjoint
representation
\be
\Phi^{\alpha \beta}(x,y) = (P exp ig \int^x_y A_{\mu}
dz_{\mu})_{\alpha \beta},~~A_\mu=A^a_\mu T^a,~~ T^a_{bc}=-if^{abc}
\ee
In computing the matrix element (11) one can integrate first over
fields $\bar A^a_{\mu}$, and then as a last step integrate over fields
$b_\mu$.
It is essential  that $\Phi$  contains both  fields
$\bar A_\mu$ and $b_\mu$
and  we shall choose the so-called modified
coordinate gauge [6] to express $  A_\mu$ in terms of field
strength $F_{\mu\nu}$. Let us now specify this gauge and to this end
choose $x,y$ in (11) on the 4-th axis, so that $x=(0,\vec
0),~y=(T,\vec 0).$
In the chosen gauge we have
\be
A_4(\vec z,z_4)=\int^1_0 d\alpha z_iF_{i4} (\alpha\vec z, z_4),
\ee
\be
A_k(\vec z,z_4)=\int^1_0\alpha d\alpha z_iF_{ik} (\alpha\vec z, z_4)
\ee
From (13) it is clear that $\Phi(0,T)\equiv 1$ and parallel
transporters can be omitted.

The basic quantity to be considered below is the gluonic propagator
$G_{\mu\nu}^{ik}$
\be
G^{ik}_{\mu\nu}(x,y)=<b^i_{\mu}(x)b_{\nu}^k(y)>_{b,\bar A},
\ee
where the averaging
over fields $b,\bar A$ should be done with the weight $exp(-\int S_E d^4x) $.
The averaging over fields $b,\bar A_{\mu}
$ will be done in two steps. First, the
averaging over  $\bar A_{\mu}$ for some operator $N$ can
be written as
\be <N(\bar A)>_{\bar A}\equiv \frac{1}{Z}\int D\bar A_{\mu} exp (-\int
d^4x S^{(0)}(\bar A) ) N(\bar A)
 \ee
 and therefore the averaging over $\bar A_{\mu}, b_{\mu}$ in
(15) implies the definition of the effective valence gluon action $S_{eff}
(b),$ \be e^{-S_{eff}(b)}=<e^{-\sum_i\int S^id^4x}>_{\bar A}
 \ee
 with $i=(1),(20),
(21), (F), (2,2), (3,1), (3), (4).$ As a next step the averaging over fields
$b$ can be  writen for some operator $T(b)$ as
\be
<T(b)>_b\equiv \frac{1}{Z}\int exp (-S_{eff} (b)) T(b) D b_{\mu}
\ee
In calculating $S_{eff}(b)$ in (17) we shall use the cluster expansion
theorem and neglect
 in the
 first approximation  all higher cumulants beyond quadratic ones
(discussion of this approximation will be given below in this
section). Then one obtains
\be
<e^{-\sum_i
S^i(b,\bar A)}>_{\bar A}
\cong e^{-\sum_i<S^i>_{\bar A}+\frac{1}{2}\sum_{i,k}<<S^iS^k>>_{\bar A}},
\ee
where  double angular brackets denote
the cumulant:
\be
\ll S^2\gg = <S^2>-<S>^2
\ee
It is easy to show that  within our approximation the terms
$S^{(2,0)}, S^{(3)}, S^{(4)}$ contribute only to the linear average
$<S^i>_{\bar A}$, while $S^{(1)}, S^{(2,1)}, S^{(F)}, S^{(3,1)}$
-- only to the
quadratic one.

Hence one has
\be
S_{eff}(b)=\int(S^{(2,0)}(b)+S^{(3)} (b) +S^{(4)}(b))d^4x+
\int<S^{(2,2)}>_{\bar A}d^4x-\sum_i L^{(i)},
\ee
where
$L^{(i)}( i=1;(2,1);( 3,1);F)$ are quadratic averages of the corresponding
terms $S^{(i)}$
 and also  the term, which will be important in what follows, is present:
$$
<S^{(2,2)}>_A=\frac{1}{2}
\chi^{ik,ab}\{b^i_{\mu}(x)b_{\mu}^k(x)<\bar A^a_{\lambda}(x)
\bar A^b_{\lambda}(x)>_{\bar A}-
$$
\be
-b_{\mu}^i(x) b^k_{\nu}(x) <\bar A^a_{\mu}(x)\bar  A^b_{\nu}(x)>_{\bar A}\}
\ee

Now using (14), (15) one can write
\be
g^2<\bar A^c_{\lambda}(x)
\bar
A^{c'}_{\lambda'}(y)>_{\bar
A}=\frac{\delta_{cc'}}{C^f_2}J_{\lambda\lambda'}(x,y),
\ee
 where
$C_2^f=\frac{N^2_c-1}{2N_c}$ and the important function
$J_{\mu\nu}(x,y)$ is introduced:
\be
J_{\lambda\lambda'}(x,y)=\int^1_0 x_i
d\alpha\alpha^{(\lambda)}\int^1_0 d\beta y_k\beta^{(\lambda')}
g^2\frac{tr_f}{N_c}<F_{i\lambda}(x\alpha)F_{k\lambda'}(y\beta)>_{\bar A}
\ee
Here $\alpha^{(\lambda)},\beta^{(\lambda)}=1$ for $\lambda =4$ and
$\alpha^{(\lambda)}=\alpha,\beta^{(\lambda)}=\beta$ for
$\lambda=1,2,3.$
Below  we shall use  $D$ and $D_1$ correlation functions defined through the
following decomposition [7]:
 \be
\frac{tr_fg^2}{N_c}<F_{i\lambda}(u)
F_{k\lambda'}(u')>_A=(\delta_{ik}\delta_{\lambda\lambda'}
-\delta_{i\lambda'}\delta_{k\lambda}) D(u,u')+
\Delta^{(1)}_{ik,\lambda\lambda'}(u,u')
\ee
with
the factor $\Delta^{(1)}$ proportional to the full derivative and
therefore not contributing to the confinement and to the string
creation; we shall disregard it below.

Now we can also rewrite (22) for $<S^{(2,2)}>$  through the kernel
$J_{\mu\nu}$:
\be
<S^{(2,2)}>_{\bar A}=\frac{N_c}{2C_2^f}
b_{\mu}^i(x)b^i_{\nu}(x)(\delta_{\mu\nu}
J_{\lambda\lambda}(x,x)-J_{\mu\nu}(x,x)).
\ee
At this point it is necessary to argue why in $S_{eff}(b)$ (21) we have
omitted the higher--order cumulants  and
shall also omit the term $L^{(2,2)}$, quadratic in $S^{(2,2)}$. The reason is
that those terms contain quartic (and higher)  terms of the type $\ll A^4\gg$
which due to (13),(14) are expressed through $\ll F^4\gg$. As it was argued
in [8] these quartic and higher cumulants do not contribute significantly to
the string tension which can be written (for the fundamental string)
as
 \be
\sigma_f=\frac{1}{2} \int\int d^2 x D(x) +0(\ll F^4\gg)
\ee
For static charges of higher SU(3) representations the correlator
$D(x)$ is proportional to the quadratic Casimir operator
$C_2$ [8] in good
agreement with lattice data [9].  The contribution $0(\ll F^4\gg)$,
which produces the $(C_2)^2$ dependence in the string tension, has
not been seen in the static potential of adjoint charges on the
lattice [9].  Additional arguments come from the string profile
calculations [10] where using only $D(x)$ one reproduces lattice data
with good accuracy.  Hence, we expect that the contribution of higher
cumulants is suppressed at least in those two instances, whereas
confinement (nonzero $\sigma$) is already present due to the Gaussian
correlator $D(x)$.

Therefore
to obtain the selfcoupled equations for the lowest
cumulants we confine ourselves to
 functions $D(x)$ and $\Delta^{(1)}$,  while at the next
stage one can express triple and quartic correlators through the
lowest ones and so on.  In this way one  gets an expansion in powers
of the parameter $\rho \equiv (g\sqrt{<F^2_{ik}>} T_g^2),$ where
indices $i,k$ are fixed and refer to the surface of integration, as
in (27).  Inserting phenomenological values of gluonic condensate
[3] and the  lattice value of $T_g$ [1], one obtains $\rho\approx
0.1$ which implies a fast convergence of cluster expansion.  Note
that the case of instantons, where the cumulant series is not
converging, requires a special treatment, which  was done e.g. in
[11].

\section{Mass operator and nonlinear equation}

The effective action $S_{eff}(b)$ in equation (21) contains terms which
are quadratic in $b(i.e. L^{(1)}, <S^{(2,2)}>, S^{(2,0)})$, and in
addition also cubic  and quartic terms  in $S^{(3)},S^{(4)}$ and
$L^{(2,1)}, L^{(F)}$, and the sixth power term $L^{(3,1)}$.
Keeping aside $S^{(3)}, S^{(4)}$
 let us turn now to other terms and write the
effective mass term $ M$ in the same way as it is usually done in the
Dyson--Schwinger formalism [12]. Writing diagrams based on the
vertices $L^{(2,1)}, L^{(F)}$ and $L^{(3,1)}$ one obtains (in the
leading order of large $N_c$) the Hartree--Fock approximation for
$ M$ generated by those vertices, which
are obtained doing  a Wick
pairing for every extra pair of operators $b_{\mu}$ with replacement
$b_{\mu}b_{\nu}\to G_{\mu\nu}$ (15).

In the same approximation the valence gluon Green's function
$G_{\mu\nu}$  (16) is diagonal in color (since the resulting mass operator
$ M$  will be shown to be is diagonal) and one can rewrite (15) as
\be
<b^\alpha_{\mu}(x)b^\beta_{\nu}(y)>_{b,\bar A}=
\delta_{\alpha\beta} G_{\mu\nu}(x,y).
\ee
As a result one can represent $M$ in the following form,
\be
M=M^{(2,2)}+M^{(2,1)}+M^{(1)}+M^{(3,1)}+ M^{(F)}+\Delta M,
\ee
where  $M^{(i)}$ correspond to $S^{(i)}$ and e.g.
\be
M^{(2,2)}_{\mu\nu}(x,y)=\frac{N_c}{C_2^f}\delta^{(4)}(x-y)
[J_{\lambda\lambda}
(x,y)\delta_{\mu\nu}-J_{\mu\nu}(x,y)],
\ee
\be
M^{(2,1)}_{\mu\nu}(x,y)=-\xi \frac{N_c}{C_2^f}J_{\lambda\lambda'}
(x,y)G_{\rho\sigma}(x,y)
N_{\lambda\rho\mu}(x)
N_{\lambda'\sigma \nu}(y)
\ee
with
$$
N_{\lambda\rho\mu}(x)
=\delta_{\mu\rho}(2\partial^G_{\lambda}+\partial^J_{\lambda})
+
\delta_{\lambda\rho}
(\partial^J_{\mu}-\partial^G_{\mu})-
\delta_{\lambda\mu}(2\partial^J_{\rho}+\partial^G_{\rho}),~~
$$
\be
\mbox{\rm and~the ~parameter~}\xi\equiv \frac{r}{N_c^2-1},
 \ee
and
$$
M^{(1)}_{\mu\nu}(x,y)=-\frac{1}{N_c^2-1}<D_\lambda^{ca}F^a_{\lambda\mu}(x)
D^{cb}_{\lambda'}
(y) F^b_{\lambda'\nu}(y)>_A=
$$
\be
=-\frac{1}{N_c^2-1}\frac{\partial}{\partial
x_{\lambda}}\frac{\partial}{\partial y_{\lambda'}}<F^c_{\lambda\mu}
(x)
F^c_{\lambda'\nu}(y)>_A+0(<FFF>)
\ee

Here in  (32) we have omitted higher
cumulant contribution and all other $M^{(i)}$ proportional to
$\xi$ will be disregarded in the large $N_c$
limit.

The terms $S^{(3)}, S^{(4)}$ generate their own contribution to
$M_{\mu\nu}$, which we  denote $\Delta M$. We stress now that all terms
in $M$ containing the valence Green's function $G_{\mu\nu}$,
acquire a factor $\xi$ for each $G$, the same also refers to $\Delta
M$. Except for $\xi$ the $N_c$ dependence in all terms in $M$
cancels and all terms are $0(N_c^0)$. Therefore
in the  first approximation one can neglect all
terms containing $\xi$ and is left with
\be
M\cong M^{(2,2)} +M^{(1)} +0(1/N^2_c)
\ee
Then Dyson--Schwinger equation for the valence gluon Green's function
can be derived from (21 ) in a usual way [12]:
\be
 (-\partial^2_\lambda \delta_{\mu\rho}
+\partial_{\mu}\partial_{\rho}) G_{\rho\nu}(x,y) +\int  M_{\mu\rho}
(x,z) G_{\rho\nu} (z,y) d^4z= \delta^{(4)}(x-y),
 \ee
  here $M$ is
given in (24-33) or at large $N_c$ approximately in (34).

Equation (35) is the central point of the paper. In the rest part of
it we analyze equation (35) in more detail and express field
correlators $D,D_1$ through $G_{\mu\nu}$, thus closing the set of
equations.

\section{ Analysis of the selfcoupled equation (43)   and  simple
estimates}

  To compute $D$ and $\Delta^{(1)}$, as defined in (25), through
  $G_{\mu\nu}$
it is convenient to choose, in each term of the sum over color indices in
(11), the color indices $\alpha, \beta$ as belonging to the small set of
valence colors. Hence fields $F_{\mu\nu}$ will consist of $b_{\mu}$ only and
(11) can be rewritten as
  $$
  \frac{tr_f g^2}{N_c} <F_{\mu\nu}(x) \Phi(x,y)
  F_{\lambda\sigma}(y)>=
  \frac{g^2}{2N_c}
  <\frac{\partial}{\partial x_{\mu}}b^i_\nu(x)\Phi^{ik}(x,y)
  \frac{\partial}{\partial y_\lambda} b_\sigma^k(y)>+
  $$
  $$
  +perm+ {gf^{cab}}<\frac{\partial}{\partial x_\mu}
  b^d_\nu (x) \Phi^{dc}(x,y) b^a_\lambda(y) b^b_\sigma (y)>+perm +
  $$
  \be
  +{g^2 f^{cde}f^{cab}}<b^d_{\mu}(x) b^e_\nu(x)
   b^a_\lambda (y) b^b_\sigma (y)>\}
  \equiv \frac{(N^2_c-1) g^2}{2N_c}\{I_1+I_2+I_3\}
  \ee
  By perm in (36) we denote the sum of terms which obtain from the
  preceding one using the rule
  \be
  (\mu\nu\lambda\sigma)+perm
  =+(\mu\nu\lambda\sigma)-(\nu\mu\lambda\sigma)
  -(\mu\nu\sigma\lambda)+(
  \nu\mu\sigma\lambda)
 \ee
   In (36) it is also assumed that finally one can put $x=(0\vec 0),
  y=(T,\vec 0)$ and therefore $\Phi(x,y)$ in the gauge (13) can be
  replaced by unity.

  The r.h.s. of (36) can be written symbolically as
$$\partial_{\mu}\partial_\nu G_{\nu\sigma}(x,y) + G_{\mu\nu} (x,y)
G_{\nu\sigma} (x,y) + perm.+ 0(<FFF>) +0 (g^2).$$
 Since the
asymptotics of $G(x,y) $ at large $|x_4-y_4| $ ( as is seen from
the integral equation  (35)  with the kernel (34))
is exponential, the same is true for the l.h.s. of (36), i.e. for
$D(x)$ and $\Delta^{(1)}$. This fact is in agreement with
lattice data [1].

The equation (35) and the kernel  $M$ do not contain dimensional
parameters and are scale--invariant. Therefore going from large
distances, where nonperturbative contributions are dominant, the
scale appears spontaneously in the nonperturbative solution. All
other nonperturbative paramerters, like gluonic condensate $\sim
(D(0)+D_1(0))$ are expressed through this prescribed scale parameter.

If one, however, starts from small distances and develops the
perturbative series like
\be
G_{\mu\nu}(x,y)= G^{(0)} - G^{(0)} MG^{(0)}+G^{(0)} MG^{(0)}
MG^{(0)}-...
\ee
 where $G^{(0)}$ is the free solution of (35), then the perturbative
 divergencies require regularization and the renormalization, which
 introduce their own perturbative scale -- normalization mass $\mu$
 or $\Lambda_{QCD}$ .
 Fixing this scale, one can extrapolate to larger distances and find
 nonperturbative solution defined at this scale. We shall discuss
 this procedure in detail in a subsequent publication,  and now
 look at the properties of the pure nonperturbative solutions.

 We shall be interested in  the large time behaviour of  gluon
 Green's function and to make an estimate we first keep in the kernel
 $M$ (34) only the local part $M^{(2,2)}$, since it grows at large
 distances,  while $M^{(1)}$ tends to a constant.

 We assume
at this point (and later confirm it in calculations) that
 $D(u)$ has an exponential form
 \be
 D(u) =D_0 exp (-\omega |u|), ~~\omega=1/T_g.
 \ee
 From (27) one has $\sigma = \frac{\pi D_0}{\omega^2}
 $ and $J_{\lambda\lambda}$ is
 \be
 J_{\lambda\lambda}(x,x) =\frac{2\sigma x\omega}{\pi}
 ((1)_E+(\frac{2}{3})_M)
 \ee
 We have used in (40) the subscripts $E,M$ to specify the
 contribution from colorelectric and colormagnetic fields
 respectively. The latter is gauge-- dependent if one keeps only
 Gaussian correlators and neglects all higher--order ones.

 One can
diminish this gauge dependence taking into account that
the inclusion of all correlators creates for every  Wilson loop the surface
 of the minimal area. In our case of the large $T$ asymptotics this  surface,
 built on the trajectory of the valence gluon and the parallel transporter
 along the time axis, is time--like.
  This means that (unless large  angular momenta $J$ are considered)
 the spacial projections $\sum_{ij}$ of the minimal surface are not
 growing with time $T$, and therefore \underline{magnetic}
  contributions to
 the linear confinement term in (40) should be cancelled in the sum of all
 correlators.
  Hence, the choosing  of the minimal area surface in the path--integral
 formalism [13] is equivalent to the temporal surface gauge [14] in
 the formalism  presented here and in [4]. As a practical outcome
  the magnetic term $(\frac{2}{3})_M$
 in (40) vanishes in the temporal surface gauge.

 From the structure of (35) one can deduce that $G_{\mu\nu}(x,y) $
 can be expanded in a complete set of  transverse gluon states

 \be
 G_{\mu\nu} (x,y) = \sum_n c_n^{\mu\nu} u_n(x)  u^+_n(y)
 \ee
 where eigenvectors $u_n(x)$ satisfy an equation
 \be
 -\partial^2_\lambda u_n(x) +\frac{N_c}{C_2^f}
 J_{\lambda\lambda}(x,x) u_n(cx) =0.
 \ee
 Writing  $u_n(x)$ as
 \be
 u_n(x) = exp (-\omega_n x_4 ) \varphi_n(\vec x)
 \ee
 one
 obtains equation for $\varphi_n(\vec x)$
 \be
 \omega^2_n \varphi_n =(-\frac{\partial^2}{\partial \vec x^2} + \bar
 C|\vec x|)\varphi_n
 \ee
 with
 \be
 \bar C=\frac{N_c}{C_2^f}\frac{2\sigma \omega}{\pi}
 \ee

 The analysis of the equation (44) was done in [15] with the result
 \be
 \omega^2_n= a_n (\bar C)^{2/3},
 \ee
  where $a_n$ are numerically computed numbers, for the lowest state
  $n=0$ one has $a_0=2.33...$

  At this point it is important to use the consequence of
  selfcoupling in (35) and (36).
  Namely, as follows from (41) and (43), the propagator $G_{\mu\nu}$
  decays exponentially at large time $T$.
   As a consequence of (36) and the analysis done above, the
  function $D(T)$ decays at large $T$ with the {\bf same exponent},
  i.e.
  \be
  \omega =\omega_0
  \ee
  Inserting (47) into (46) one obtains
  \be
  \omega =(2.33)^{3/4} \sqrt{\frac{9\sigma}{2\pi}}
  \ee
  and  with $\sigma = 0.2 GeV^2$, the final result is
  \be
  \omega=1/T_g = 1.01 GeV
  \ee
  This value agrees  well with the lattice measured  value
  $T_g=0.22 fm (\omega = 0.91 GeV)$ [1].

   Thus we have demonstrated in this example that different
   parameters of  the nonperturbative correlators can be expressed
   through each other via a dynamical computation.

   \section{Conclusions}

The analysis of the previous sections was intended to show that
selfcoupled equations (30), (34-36) indeed contain a nontrivial
dynamics, which allows to obtain nonperturbative gluon propagators
and correlators.
Numerical solution of these equations is possible and is now in
 progress, in this paper however we confine ourselves to simple
estimates and qualitative analysis. The characteristic behaviour of
the gluon propagator $G(x_4,y_4,\vec x, \vec y)$, which one finds
from (35) and (43), is exponential decay in $|x_4-y_4| $ and the
faster than exponential decay at large $|\vec x|$  or $|\vec y|$. The
latter property signals the  confinement of the gluon, since the
adjoint string at large $N_c$ does not allow the gluon to go far from
the origin. The parallel transporter along the temporal axis, present
in our definition of the gauge--invariant gluon propagator, is
actually a signal of the static adjoint charge at the origin,
therefore the gauge invariant quantity $G_{\mu\nu}$ is in fact the
propagator of the white system made of the gluon and the static
charge. Hence  the mass $\omega$  in the exponential decay of
$G_{\mu\nu}$ in time  is equal to the mass of this white system.

Connection between $G_{\mu\nu}(x_4,y_4, \vec 0\vec 0)$ and
$D(x_4-y_4)$ and $D_1(x_4-y_4)$ found in section 4, tells us
immediately that $D$ and $D_1$ decay with the same dominant exponent
with mass $\omega$.

In the subsequent  paper  we shall discuss the perturbative
contents of our equations (35) and  of the effective action $S_{eff}(b)$
(21).

One of the most important problems to be solved in the framework of
the suggested formalism is the problem of the deconfining phase
transition. The latter may occur or due to the increase of
temperature or due to the addition of scalars with the Higgs--type
interaction. In the last case one can get a system of equations of
the type of (35) for the gluon and scalar propagators which describe
two possible phases: confinement and deconfined massive gluons. We
plan to analyze these questions in forthcoming papers.

The author is grateful  to A.Di Giacomo, H.G.Dosch,
 M.Schmidt and V.I.Shev\-chen\-ko for useful discussions and to
 A.M.Badalian for valuable remarks.

This work was done with the partial support of grants  INTAS
94-2851, 93-79 and RFFI grant 97-02-16404.

  \end{document}